\documentclass{iau-JDSS}
\usepackage{graphicx,bm,url}

\newcommand{\EQ}{\begin{equation}}
\newcommand{\EN}{\end{equation}}
\newcommand{\EQA}{\begin{eqnarray}}
\newcommand{\ENA}{\end{eqnarray}}

\newcommand{\Fig}[1]{Figure~\ref{#1}}

{}
{}
{}
\newcommand{\meanemf}{\overline{\cal E} {}}

{}
{}
\newcommand{\meanEMF}{\overline{\mbox{\boldmath ${\cal E}$}}{}}{}
{}
{}
{}
{}
{}
\newcommand{\meanBB}{\overline{\mbox{\boldmath $B$}}{}}{}
{}
{}
{}
{}
{}
{}
{}
\newcommand{\meanJJ}{\overline{\mbox{\boldmath $J$}}{}}{}
\newcommand{\meanUU}{\overline{\bm{U}}}

{}
{}
{}

\newcommand{\meanB}{\overline{B}}

\newcommand{\meanJ}{\overline{J}}

{}

{}
{}

\newcommand{\xx}{\bm{x}}

\newcommand{\ff}{\mbox{\boldmath $f$} {}}

\newcommand{\nab}{\mbox{\boldmath $\nabla$} {}}

\def\Pm{P_{\rm m}}
\def\Rm{R_{\rm m}}

\def\cs{c_{\rm s}}

\def\kf{k_{\rm f}}

\def\urms{u_{\rm rms}}

\def\etat{\eta_{\rm t}}
\def\etatz{\eta_{\rm t0}}

\newcommand{\yapj}[3]{ #1, \textit{ApJ,} \textit{#2}, #3}

\newcommand{\yan}[3]{ #1, \textit{Astron.\ Nachr.,} \textit{#2}, #3}

\newcommand{\ymn}[3]{ #1, \textit{MNRAS,} \textit{#2}, #3}

\newcommand{\ybook}[3]{ #1, \textit{#2} (#3)}

\title[Turbulent diffusion]
{Turbulent diffusion and galactic magnetism}

\author[A. Brandenburg \& F. Del Sordo]
{Axel Brandenburg \& Fabio Del Sordo}

\affiliation{NORDITA, Roslagstullsbacken 23, SE-10691 Stockholm, Sweden; and\\
Department of Astronomy, Stockholm University, SE-10691 Stockholm, Sweden}

\pubyear{2009}
\volume{Volume 15}  
\pagerange{119--126}
\date{?? and in revised form ??}
\jname{Highlights of Astronomy, Volume 14}
\editors{Ian F Corbett, ed.}
\begin{document}

\maketitle

\begin{abstract}
Using the test-field method for nearly irrotational turbulence driven
by spherical expansion waves it is shown that the turbulent magnetic
diffusivity increases with magnetic Reynolds numbers.
Its value levels off at several times the rms velocity of the turbulence
multiplied by the typical radius of the expansion waves.
This result is discussed in the context of the galactic mean-field dynamo.
\keywords{turbulence, (magnetohydrodynamics:) MHD, galaxies: magnetic fields}
\end{abstract}

The galactic dynamo is believed to be powered by supernova-driven turbulence.
This type of forcing does not directly produce vorticity; it can
only be produced indirectly through oblique shocks, i.e.\ through
the baroclinic term.
The aim of this work is to assess whether vorticity is actually important
for the dynamo.

The galactic magnetic field has a strong large-scale component which is
generally believed to be due to a mean-field dynamo of $\alpha\Omega$ type
that is governed by the equation
\EQ
{\partial\meanBB\over\partial t}=\nab\times(\meanUU\times\meanBB
+\meanEMF-\eta\mu_0\meanJJ),\quad\mbox{where}\quad
\meanemf_i=\alpha_{ij}\meanB_j-\eta_{ij}\mu_0\meanJ_j
\EN
is the mean electromotive force, $\meanJJ=\nab\times\meanBB/\mu_0$
is the mean current density, $\meanUU$ is the mean flow,
and $\mu_0$ is the vacuum permeability.
In order to assess its efficiency, one needs to determine the tensors
$\alpha_{ij}$ (the ``$\alpha$ effect'') and $\eta_{ij}$ (turbulent
magnetic diffusivity).
Note that $\alpha_{ij}$ is a pseudo tensor and non-vanishing diagonal
components can only be constructed from a combination of polar and axial
vectors, and would therefore be vanishing in the absence of stratification
and rotation.
The $\eta_{ij}$ tensor, on the other hand, does not require this, and
it should be finite even in the completely homogeneous case.
This is the case considered in the present study, which is a
necessary intermediate step.

For homogeneous flows $\eta_{ij}$ is an isotropic tensor,
which we write as $\eta_{ij}=\etat\delta_{ij}$, where $\etat$ is
the turbulent magnetic diffusivity.
A relevant concern in mean-field theory is that turbulent transport
coefficients such as $\etat$ must stay finite even in the limit
of large values of the magnetic Reynolds number, defined here
as $\Rm=\urms/\eta\kf$, where $\urms$ is the rms velocity of the
turbulence, and $\kf$ is the wavenumber corresponding to the scale of
the energy-carrying motions.
Given the importance of a possible $\Rm$ dependence, it is necessary
to perform so-called {\it direct} simulations, where no subgrid scale
modeling is used.
This implies that we must make compromises regarding the strength of
the forcing and consider only subsonic flows.
Following earlier work of \cite{MB06} we consider a flow
driven by random expansion waves of radius $R=2/\kf$ (not to be confused
with the magnetic Reynolds number $\Rm$) and determine $\etat$
using the test-field method of \cite{Sch05} in the implementation of
\cite{B05_QPO}.

The evolution of internal energy and hence entropy is not relevant
to our question about turbulent transport coefficients.
Therefore we consider an isothermal equation of state where the
pressure $p$ is proportional to the density $\rho$ with
$p=\rho\cs^2$, with $\cs$ being the isothermal speed of sound.
We adopt a Gaussian potential forcing function $\ff$ of the form
$\ff(\xx,t)=\nab\phi$ with
$\phi(\xx,t)=N\exp\left\{[\xx-\xx_{\rm f}(t)]^2/R^2\right\}$,
where $\xx=(x,y,z)$ is the position vector,
$\xx_{\rm f}(t)$ is the random forcing position,
$R$ is the radius of the Gaussian, and $N$ is a normalization factor.
We consider a time dependence of $\xx_{\rm f}$ with a forcing time
${\delta{}t}_{\rm force}\approx(\urms\kf)^{-1}$ that defines the
interval during which $\xx_{\rm f}$ remains constant.
We use the \textsc{Pencil Code} (\url{http://pencil-code.googlecode.com})
which is a non-conservative, high-order, finite-difference code (sixth
order in space and third order in time) for solving the compressible
hydrodynamic and hydromagnetic equations.

\begin{figure}[t!]\begin{center}
\includegraphics[width=.5\columnwidth]{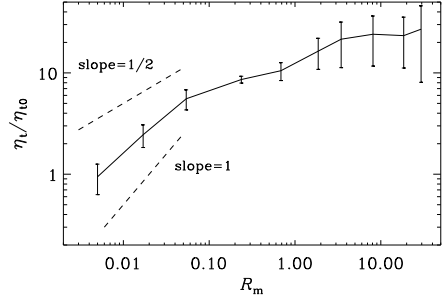}
\end{center}\caption[]{
Dependence of $\etat$ on $\Pm=1$.
}\label{peta_Rm_PrM1}\end{figure}

In \Fig{peta_Rm_PrM1} we plot the dependence of $\etat$ on $\Rm$.
Following earlier work of \cite{Sur08} we normalize $\etat$ by
$\etatz\equiv\urms/3\kf$.
Note that, for low values of $\Rm$, $\etat$ increases
proportional to $\Rm^n$ with $n$ between 1/2 and 1.
For larger value of $\Rm$, $\etat$ seems to levels off at a value of about
20 times $\etatz$.
Expressing this in terms of $\urms$ and the typical radius $R$ of the
expansion waves, we find that $\etat\approx4\urms R$.
Note also that $\etat$ is always positive, in contrast to analytic
predictions for irrotational turbulence using the first-order smoothing
approximation \cite{KR80}.

Based on these results we can conclude that nearly irrotational turbulence
is at least as efficient as vortical turbulence in diffusing mean
magnetic field.
Clearly, our study is still at a preliminary stage.
It is important to clarify a possible dependence of our results
on the microscopic magnetic Prandtl number and, in the nonlinear
regime, on the magnetic field strength.
Next, we need to consider the case with rotation and stratification
which should then lead to an $\alpha$ effect, as well as turbulent pumping.
This would provide an opportunity to compare with early
predictions by \cite{Fer92} for this type of flows.

\begin{acknowledgments}
We acknowledge the allocation of computing resources provided by the
Swedish National Allocations Committee at the Center for
Parallel Computers at the Royal Institute of Technology in
Stockholm and the National Supercomputer Centers in Link\"oping.
This work was supported in part by
the European Research Council under the AstroDyn Research Project 227952
and the Swedish Research Council grant 621-2007-4064.
\end{acknowledgments}

\end{document}